# Using SEQUAL for Identifying Requirements to Ecore Editors

Kristian Rekstad and John Krogstie
John.Krogstie@ntnu.no
Department of Computer Science, NTNU, Trondheim Norway

**Abstract:** Software engineers who use Model-Driven Development may be using Ecore for their work. Ecore is traditionally edited in Eclipse IDE, but a recent transition to Web tools allows for development of new Ecore editors. To investigate the needed functionality of such modeling tools, the model quality framework SEQUAL has been applied. The paper presents the current results of this task, producing requirements for tool functionality as quality improving means for the following quality aspects: physical, empirical, syntactic, semantic, pragmatic, social and deontic. The result is an extensive list of tool functionality that could be implemented by the Ecore editor developers. Although many requirements are identified, the framework should also help in making trade-offs in case not all requirements can be implemented. In this way the paper both contribute to identifying modeling tool functionality, and to have input to improve SEQUAL as a general model quality framework. Further work will need to be done on the implementation of the tools for properly evaluating this work.

## 1 Introduction

In the field of software engineering, one approach to the development of software is through modeling. Modeling is the act of abstracting a phenomenon in a domain into a model in a purposeful way [24], such as a diagram of entities with properties and relations. There are various degrees of how the model controls the software: from being pure documentation in Model-Based Development, to being the source of (more or less complete) code generation in Model-Driven Development (MDD), or to being the actual target of execution itself [3].

Within the field of Model-Driven Development, the goal is often to derive most of the software from a set of models [25]. Details that can't be modeled can be programmed manually after code generation. One such tool for creating models and generating code is the Eclipse Modeling Framework (EMF). It allows modelers to express their domain using a meta-model called Ecore (from EMF core). After modeling, EMF can augment the model with a genmodel file ("Generator model"). The genmodel allows a programmer to enter any details related to java code generation, such as package name and choice of code generation templates.

The Ecore meta-model is an implementation of Essential Meta Object Facility (EMOF) in the OMG Meta Object Facility (MOF) Core Specification. This makes Ecore analogous to UML Class Diagrams, to some extent. [17] Ecore is object-oriented, and the primary entity is classes (EClass) with attributes (EAttribute and EReference). However, a developer is free to create metamodels with other entities using Ecore; the framework has no restriction on whether the user makes models, meta-models or meta-metamodels etc., being appropriate for supporting domain specific modeling (DSM [6]). In a typical case, Ecore with be level M2 in a model hierarchy ([10], p. 101), and Ecore is its own metamodel. [17] The entities in Ecore are shown in Figure 1. EClass, EDataType and EEnum are the main entities. Inside an EClass there can be: EAttribute, EReference, EOperation and EAnnotation. Inheritance is done in an EClass using an EReference named eSuperTypes (with no upper bound, allowing multiple-inheritance).

The traditional approach to authoring models in Ecore has been through the Eclipse IDE with EMF plugins. The available tools for editing models are:
- a text editor for XML (XMI)
- a master-detail view with a tree and property sheet
- a diagram editor like an editor for UML diagrams.

Diagram editors are being moved to the web. Eclipse is a desktop application, and the diagram editor is created by a framework called Sirius. The developers behind Sirius are working on Sirius Web, which deploys editors to web browsers instead of Eclipse. [2]. Likewise, the developers behind EMF are working on plugins for Theia. Theia is an IDE by the Eclipse Foundation that aims to be a platform like the Eclipse IDE. A

```
EAttribute -> EStructuralFeature
EAnnotation -> EModelElement
EClass -> EClassifier
EClassifier -> ENamedElement
EDataType -> EClassifier
EEnum -> EDataType
EEnumLiteral -> ENamedElement
EFactory -> EModelElement
EModelElement
ENamedElement -> EModelElement
EObject
EOperation -> ETypedElement
EPackage -> ENamedElement
EParameter -> ETypedElement
EReference -> EStructuralFeature
EStructuralFeature -> ETypedElement
ETypedElement -> ENamedElement
EGenericType
ETypeParameter -> ENamedElement
```

Figure 1: All the main concepts in the Ecore metamodel. What is left out are EDataTypes (EInt, EString etc.) and EStringToStringMapEntry. The left name is entity name, the arrow shows inheritance, the right name is eSuperTypes (parent class being inherited).

project called ecore-glsp is currently in progress, which provides a graphical diagram editor in a browser, using Theia [4].

Both new projects (Sirius Web and ecore-glsp) are good opportunities to apply practices for ensuring good model quality, because the tools themselves are not developed yet. Thus, one can shape the tools by supporting functionality as means to develop good models. Additionally, this is a good moment to specify requirements for the tools, so that users will have an easier time creating models of high quality when using the tools.

This paper reports parts of a larger design science research effort on developing appropriate web-based tools for supporting MDD. We assume it helps to have a framework for model quality to discuss support for developing models of high quality through the application of Ecore tools. The framework chosen in this work is the Semiotic Quality Framework - SEQUAL, defined in [10,13]. Although other frameworks for quality of models exist e.g. [1,19,10] SEQUAL has some features that makes it specifically appropriate for this task. In particular it distinguishes between quality characteristics (goals) and means to potentially achieve these goals by separating what you are trying to achieve from how to achieve it. One class of means is to have appropriate tool support for developing models in general of high quality.

The research questions for this part of the work are

- RQ1 Is it helpful to use SEQUAL to develop the requirements for an MDD-modeling tool?
- RQ2 How can the use of SEQUAL in this case contribute to improve SEQUAL itself, relative to support the development of modeling tool requirements?

Section 2 briefly introduce SEQUAL. Then section 3 will attempt to specialize SEQUAL for requirements for tools to help develop models of high quality in Ecore model editors. Section 4 provide a discussion and conclusion.

## 2    Introduction to SEQUAL

SEQUAL [10] is a framework for assessing and understanding the quality of models and modelling languages. The framework has been developed through several iterations and have also in some cases been established as part of the knowledge base e.g., in the development of a framework for quality of maps [21]. The framework has been used for evaluation of modelling and modelling languages from many perspectives, including data models [11], and models in MDE [14], process models [13], interactive models [16], enterprise models [15], requirements models [7, 8] and goal models [9]. It has been used both for models on the type level and instance level (i.e., data quality [12]).

As mentioned above, a part of the framework that has not been used equally often relates to requirements for tools to support the *development* of good models in a selected modeling language. This aspect is something that we do not find in other model-quality frameworks, which focus on model or language quality [18,19,20, 23, 27, 28].

A benefit of using SEQUAL is that it is comprehensive, covering several different aspects in an integrated manner, including quality of models, modeling languages, modeling tool functionality, and modelling method characteristics. Another benefit is that the framework is general, allowing it to be applied to new models and modeling languages, unlike modeling language or modeling-perspective-specific quality guides. A third benefit is that new specializations can benefit from existing specializations by analogy. This specialization for example, takes inspiration from specializations of SEQUAL for Data Models (in [10,11]) and Model-Driven Software Engineering (in [14]).

Quality has been defined referring to the correspondence between statements belonging to the following sets:

- **$G$,** the set of goals of the modelling task.
- **$L$,** the language extension.
- **$D$**, the domain, i.e., the set of all statements that can be stated about the situation. Domains can be divided into two parts, exemplified by looking at a software requirements specification model:
  - Everything the computerized information system (CIS) is supposed to do. This is termed the *primary domain*.
  - Constraints on the model because of earlier baselined models such as system level requirements specifications, enterprise architecture models, statements of work, and earlier versions of the requirement specification. This is termed the *modelling context*.
- **$M$**, the externalized model itself.
- **$A$,** the part of the model that a given social or technical actor has access to
- **$K$**, the explicit knowledge of the audience (social and technical actors) relevant to the domain **$D$**.
- **$I$**, the social actor (human) interpretation of the model
- **$T$**, the technical actor (tool) interpretation of the model

    The main quality types are:
- Physical quality: The basic quality goal is that the externalized model **$M$** is available to the relevant social and technical actors (and not to others) for interpretation (**$I$ and $T$**) and further development and use to achieve **$G$**.
- Empirical quality deals with comprehension and predictable error frequencies when a model **$M$** is read by different social actors (persons)
- Syntactic quality is the correspondence between the model **$M$** and the language extension **$L$**.
- Semantic quality is the correspondence between the model **$M$** and the domain **$D$**.

- Perceived semantic quality is the similar correspondence between the social actor interpretation *I* of a model *M* and his or her current knowledge *K* of domain *D*.
- Pragmatic quality is the correspondence between the model *M* and the actor interpretation *(I* and *T)* of it.
- The goal defined for social quality is agreement among actor's interpretations.
- The deontic quality of the model relates to that all statements in the model *M* contribute to fulfilling the goals of modelling *G*, and that all the goals of modelling *G* are addressed through the model *M*.

The overall framework is illustrated in Figure 2.

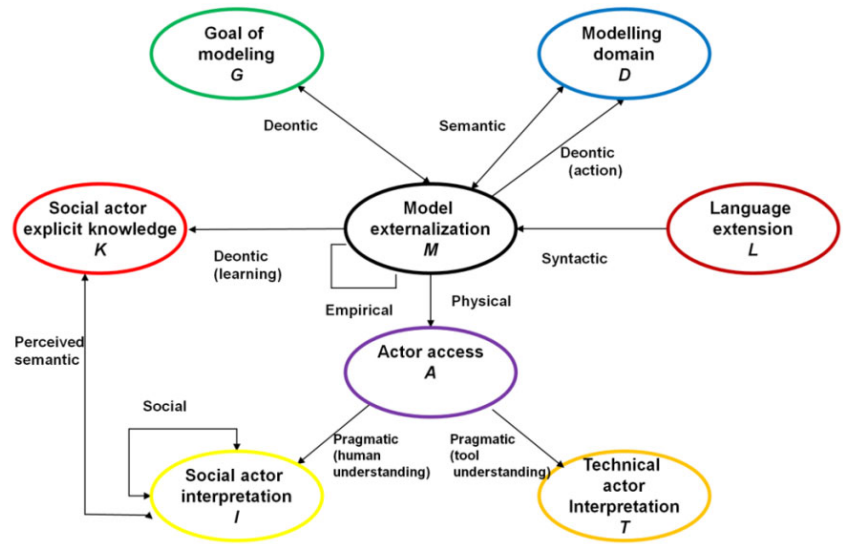

Figure 2: SEQUAL with sets and relationships between the sets (from [8])

## 3 Application of SEQUAL for identifying requirement to specialized graphical editors in EMF Ecore

We will here discuss requirement on the different quality levels in SEQUAL. First it is useful to specialize the sets in the context of MDD:

- **G,** generate executable code, or code skeletons from the models, and management of the co-evolution of models and code.
- **L,** what can be expressed in Ecore
- **D**, the domain, i.e., the set of all statements that can be stated about the situation.
  - Everything the computerized information system (CIS) is supposed to do.

- Constraints on the model because of earlier baselined models such as requirements specifications, design models and other models on a higher level of abstraction.
- *M*, the externalized model accessible in the tool.
- *I*, the main social actors in this setting are system developers.
- *T*, the technical actor (tool) interpretation of the model to do necessary code or code-skeleton generation.

The following mapping of tool-functionality to quality level was originally done by the first author, and then confirmed by the second author.

### 3.1 Physical quality

The main aspects of physical quality are persistence, currency, availability, and security. For a diagram editor, the editor framework or ecosystem encompassing the editor would be responsible for security (e.g., access control). Thus, a discussion of security (when needing to limit the access to the model of some reason) is left out here.

A common tool for versioning, code history, merging and conflict resolution used today is *git*. Some editor platforms like Gitpods (which also uses Theia) are centered around git. Therefore, some features related to git will be mentioned. The relation between git and Ecore is that the Ecore model file would be stored in a git repository.

The following bullet-points suggest a range of features which a diagram editor could implement, to achieve high physical quality for the models it edits:

- Persistence (against data loss, efficient space usage):
    - Temporarily save dirty state before the user manually saves. This avoids data loss if modeling sessions are prematurely exited/interrupted.
    - Save the diagram layout to files available for versioning (git).
    - Do not reorder file structure on save. Reordering makes for challenging git merge since it works on textual input. (iOS Xcode Storyboard merge is an example of poor practice).
    - Save on exit, or block exit to prompt the user to save first.
    - No annotations, comments or notes on the model that is not stored as files.
    - Automatic syncing (add, commit, push) to git on save. Preferably wait with pushing to git until the model serialization is syntactically validated.
- Currency (*when* something was done):
    - Display model element age on EClass and EStructuralFeature. The age could be visualized with coloring, hierarchy, or a separate table with elements and ages.

- Display changes based on git commits. (Highlight changed/added EClass, EStructuralFeature).
- Add metadata to the serialization, or a separate logfile with creation and modification of EClass/EStructuralFeature.

- Availability:

    - Serialize Ecore to standard format: XMI.
    - Possibly export to EMF-JSON. This format is more human readable while being usable for tools. A separate tool could translate between JSON and XMI if a tool demanded it.
    - Export svg/png of the displayed diagram, region, or selection. Save as file or print.
    - Share link to open the diagram in web editor, e.g., Gitpod.
    - Possible integration with external model repository for Ecore. The organization would either have to provide a model repository as a service or use a third-party service.
    - Possible to connect to a server to communicate what parts of the model are being edited now, for collaboration/coordination. (Obeo Cloud Platform for Sirius has demonstrated this in [2]).
    - Provide a shared interface for other model tools to use, like EMF.Cloud Model Server or a custom Model Server. This can provide a consistent view of models for multiple tools and send change notifications to interested parties.

### 3.2 Empirical quality

Empirical quality is about how comprehensible a model is. Some naïve examples of bad comprehensibility are: if all EClasses are stacked on top of each other, a human would not be able to understand the diagram. If EClasses had black background, and their EStructuralFeatures were written with black text, a human would not be able to read the text.

Because Ecore is close to UML Class Diagrams, most of the existing guidelines for UML Class Diagrams and different types of data modeling languages will also apply. The following list has suggestions that may improve the empirical quality:

- Avoid small angles between edges.
- Space edges so lines and labels don't overlap. (Sirius in Eclipse IDE offends this by stacking edges).
- Place EReference label and multiplicity near the correct EClass.
- Minimize the area in the diagram. (Sirius in Eclipse IDE offends this, spacing EDataTypes horizontally along the top edge before starting to lay out EClasses).
- Minimize bends and crossings of edges.
- Minimize total length of edges and longest edge.
- Place any EClass with many containment EReferences central in the diagram.

- - Alternatively place them near the top and create a tree structure.
- Order hierarchically top-down based on inheritance (eSuperTypes) or use the "dynamic instance" root EClass as tree root[1].
  - Allow user to select an EClass to order around. Sometimes, inheritance or edge count is not the right choice, and the modeler knows best what should be central.
- Enforce entered names to be PascalCase for EClass, and camelCase for eStructuralFeature, as Ecore maps to Java.
- Apply spell check to ENamedFeature name, using a user specified dictionary or English, customized to programming (e.g., splitting "myBadSpleling" into "my bad spleling" before dictionary lookup).
- Highlight invalid EClass with red border, invalid EStructuralFeature with red text or squiggly underline, and invalid EReference with red labels + edge + EClass.
- Use a different color and/or thickness for eSuperTypes edge than EReference edge.
- Use a different color for derived EReference than regular EReference.
- Some editors choose to support "dark mode" or other custom theming. If this is done, make sure the theme colors to not imply any new "meaning" that is not actually there in the model, and that the contrasts between concepts' style is preserved. Also make sure the theming does not rely heavily on colors that convert to the same grayscale or black-and-white colors.
  - Also make sure the theme colors do not use colors normally associated with errors/invalid models for valid models: e.g., red EClass background/borders for a valid EClass instance, where red usually signals errors.
- Manual arrangement of one or more EClass simultaneously. E.g., selecting 3 EClass and dragging them to the side.
- Manual arrangement of EReference, including shifting and adding/removing bends, and moving anchor points on the EClass.
- Snap EDataType and edges to a grid.
- Align EClass relative to other EClass (align top/bottom/side). Either during dragging an EClass or triggered with a button.
- Display that an EModelElement has EAnnotations applied to it (related to language quality). The annotations can specify constraints and validation rules using OCL, and it may therefore be useful to know if an entity has annotations.

---

[1]Eclipse Modeling Framework has named the model instances "dynamic instance" in the Eclipse IDE plugin.

### 3.3 Syntactic quality

Most of the quality issues should be ensured through the use of palettes that constrain the diagram to only have valid entities. This avoids syntactic invalidity.

- Error prevention can further stop users from creating invalid diagrams. Some goals can be:

  - Immediately warn user when entering invalid name (e.g., spaces and symbols or empty string) into ENamedFeature name
  - Prevent invalid values for EReference lowerBound and upperBound
  - Some EStructuralFeatures have fields of free text. For fields that reference data types (like EType) that are EClassifier (EDataType, EClass etc), a chooser dialogue or drop-down with valid entries is beneficial.

- Error detection:

  - EDataTypes with non-existing instanceTypeName. The generated code will not compile.
  - Validation that allows editing in an invalid (temporary) state.
  - Syntactic incompleteness checks for required properties.
  - Enforce unique names for EStructuralFeature, also across inherited features.
  - Enforce unique EClass names inside EPackage

Much work on e.g., UML Class diagrams for defining conceptual schemas has been done that is also relevant on the syntactic level here. Aguilera et al. [1] for instance found more than 100 potential issues only in class diagrams. We do not re-iterate these here.

### 3.4 Semantic quality

This quality type relates the statements in the domain **D** to the externalized model **M**. The quality issues are validity or completeness of the model. A valid model has no statements outside of the domain, and no incorrect statements about the domain [10]. E.g., consistency checks of the model can detect semantic issues.

- Warnings when the model has satisfiability/instantiation issues:
  - Warn when abstract EClass is not inherited.
  - Warn when EReference has changeable set to false, lower bound set to 1 and no default value. (The generated code can only return null, which is pointless. And a valid model instance XMI cannot be created.)
  - Error when the eSuperTypes creates circular inheritance, demonstrated in Figure 3.
- Support "dynamic instance" instantiation to enter domain object data, to test the instantiability of the model.

- A graphical editor would provide functionality to select the EClass to create a new instance from. This is a separate file created from an EClass selected as a container/root node, where the modelers fill out an XMI. This *may* need a separate editor from the graphical editor (which could be graphical as well) and may require cooperation with the graphical editor to handle renaming and other model changes.

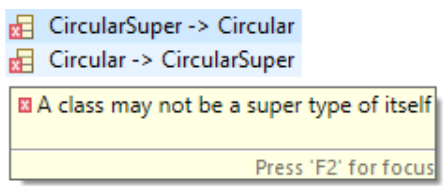

Figure 3: Circular inheritance is invalid.

Because the domain **D** is mainly accessible through the modelers' knowledge **K**, and compared to their interpretation **I** of the model, it makes sense to have quality goals supporting this comparison. The quality types of perceived semantic quality are perceived validity and perceived completeness. Quality is improved by adding or removing statements in **M** as statements are considered lacking or invalid respectively.

- Aid in comparison of **I** and **K** by reducing the number of statements in **M** that needs interpretation:
  - Generate a changelog of model insertions and deletions between two versions (e.g., git hashes). This changelog could be provided in a code review or model consolidation meeting to focus on recent changes. This would save effort in comparing previously accepted statements of **M**, by narrowing **I** to a subset.
  - Modelers could mark EClass or EStructuralFeature as "resolved" or "valid", meaning they were evaluated and considered valid. This would allow the statement checking by a social actor to skip previously evaluated statements. (This "mark" is metadata that may need persisting in a file, for physical quality).
- Tool-tips for different elements can provide clarity for what they are, improving interpretations in **I**. One example is an explanation for the *unique*, *derived* and *container* properties of EReference, which could help a social actor to correctly interpret what those values mean, for their own interpretation **I**.
  - Tool-tips are also helpful for training participants, related to pragmatic quality.
- Warnings for an EClass with no EStructuralFeatures, as this empty entity is likely a completeness issue.

### 3.5 Pragmatic quality

Pragmatic quality is about model comprehension. Two types of actors need to comprehend the model **M**: social actors (humans) and tools. Their interpretations of the model are denoted **I** and **T** for social actor and tool actor respectively. Tool interpretation should be possible as long as the model is syntactically correct, and the serialization uses an accepted format (like XMI) and semantic quality as for completeness and validity is handled to be able to generate code from the models. Thus, the focus here is mostly on functionality needed to aid human comprehension:

- Model navigation with zoom and pan.
- Show tool-tips/help text for selected entities, especially attributes on EClass and EStructuralFeature. Possibly link to a guide or full documentation for the entities when possible.
- Model filtering. "Isolate" means to hide other entities.
   - Isolate selection of EClasses and hide selection.
   - Hide edge labels or multiplicities.
   - Isolate an EClass and its eSuperTypes.
   - Isolate an EClass and its sub-classes (where the sub classes have the selected EClass as eSuperTypes).
   - Isolate an EClass and EClasses related by its EReferences.
   - Hide EStructuralFeature.
   - Isolate/hide based on query (e.g., OCL or entity type: EClass, EEnum, EDataType).
- Trigger code generation (e.g., via genmodel) on demand. The generated code can clear up confusion related to concepts in Ecore if the modelers question the impact of a model statement.
- Show information about an entity (e.g., based on git log), with *who* edited/added and entity and *why*.
- Free-text fields for providing documentation about entities. This can capture the *why* behind a model statement.
   - (This text could be added as an EAnnotation with *http://www.eclipse.org/emf/2002/GenModel* as source, and a key *documentation*. The documentation text will then carry over into the generated code).
   - The text must be available from the diagram, e.g., via a dialogue/popup or as an entry in a properties/details panel.

### 3.6 Social quality

Social quality is achieved by having agreement between social actors, with regards to the model statements. They can agree in their interpretations **I**, their knowledge **K,** or the model **M**. A modeling tool could help the processes used for achieving agreement, mainly through supporting comparison of models and merging.

- Provide views for git merge and git diff. This could be useful for code review. (E.g., using EMF Compare).
- Side-by-side comparison of models.
  - Applying automatic model-matching (e.g., using EMF Compare) to hide equal entities or highlight differences. One can use signatures or name equality; Ecore does not have unique identifiers[2].
- Copy-paste entities (EClass, EStructuralFeature etc.) across models.
- Import an entire model into another. (Verify that no EClassifier with the same name exists inside the same EPackage before importing. Or let modelers resolve name conflicts afterwards using delete/merge/rename functionality.).
- Pattern- or text-based search-and-replace for names across EClassifier sub-classes or EStructuralFeatures. This helps resolve naming issues where two modelers used synonyms and want to be consistent with regards to a glossary. E.g., replace all occurrences of "Human" with "Person" in a model.
  - Common functionality for search-and-replace is case sensitivity and regex. It could also be beneficial to select scope: which concepts should be searched (e.g., EClass vs EStructuralFeature) and which attributes (e.g., name vs defaultValueLiteral).
  - Another related option is connecting a spell checker to a glossary, to avoid usage of words outside names agreed upon in the domain . However, modeling can be the activity that discovers some of these names in the first place.

### 3.7 Deontic quality

Deontic quality is about achieving the goals of the organization, impacting the modeled domain, and teaching social actors about the domain. For Ecore, the goal is usually working software at a reduced cost, and up-to-date documentation in the form of diagrams. Ecore can especially reduce costs when the target software should have data editors for model instance creation/manipulation, such as forms and fields for entering domain data (e.g., filling in a movie database or person registry). This is because the genmodel is pre-configured with model-to-text transformations which produce java-based editors. Other transformations can be found in the community or created by the organization's developers (but at an increased cost).

A diagram editor could mainly aid learning, by exporting diagrams as documentation, or entity documentation as text/pdf/html etc. (as discussed in section 3.5 with free-text genmodel documentation). Such documentation could help increase social actor knowledge **K**.

---

[2]EClassifier has getClassifierID(), but it only works after initialization of a static model, not for a dynamic EClassifier. (17] p. 134) The generated source code reveals it as a simple indexOf on the EPackage classifiers.

## 4 Conclusion

Basing requirements for a graphical editor for Ecore diagrams using the SEQUAL Framework yields several results. Returning to the RQs we have

- RQ1 Is it helpful to use SEQUAL to develop the requirements for an MDD-modeling tool?

The list of quality measures is not necessarily exhaustive, and derives from analogy to similar SEQUAL specializations, prior work in language quality, and identified need and experiences from modeling tool design. Applying the framework allows to cover the main aspects that affect model quality, in a structured manner. SEQUAL can therefore be valuable to tool developers to create high quality tools that support modelers to produce high quality models.

A limitation is that the identified features is now a long list with little internal structure and ways of prioritization. We have seen on work with model quality that certain e.g., syntactic errors are more important than other for pragmatic quality and ultimate value (deontic quality) of the model [10]. One has also identified tradeoff between different quality levels, e.g., improving completeness as part of semantic quality might hamper pragmatic quality. Similar issues of trade-off can be relevant when discussing tool-functionality. On the other hand, what is important in different projects might vary, thus, to have a general tool that are to support many different projects it might be hard to predict all functionality that is useful. That said, in an agile fashion, the best strategy is to implement a limited set of functionalities, have other features in a product queue where prioritization on functionality in future releases is based on experience in the use of the tool, using good practice from software products research [22].

A more comprehensive evaluation of the usefulness of this approach will first be possible to make when a running version of the tool is developed.

- RQ2 How can the use of SEQUAL in this case contribute to improve SEQUAL itself, relative to support the development of modeling tool requirements?

Several the more specialized ideas described in this paper can be generalized to modelling functionality that is useful for modelling tools in general. For instance, techniques for model comparison and model validation described in this paper can be generalized to be part of the general SEQUAL framework. Another improvement is to link certain tool-functionality to several quality types, when it contributes to several types (e.g., how tool-tips can contribute both to semantic and pragmatic quality)

Further work on the approach relates to implementing the tool and getting practical experience from the use of tools to evaluate if the implemented functionality is in fact useful for producing high quality models. Another task is to take functionality identified for this type of modeling and update the more general aspects of tool quality in the SEQUAL framework [10].